\documentclass[twocolumn,twoside,slac]{revtex4}
\usepackage{amsmath}
\usepackage{graphicx}
\usepackage{fancyhdr}
\begin{document}

\title{Event Selection Using an Extended Fisher Discriminant Method}

%

\author{Byron P. Roe}
\affiliation{University of Michigan, Ann Arbor, Michigan 48109, USA}

\begin{abstract}
This note discusses the problem of choosing between hypotheses in a 
situation with
many, correlated non-normal variables.  A new method is introduced
to shrink the many variables into a smaller subset of variables
with zero mean, unit variance, and zero correlation coefficient
between variables.  These new variables are well suited to use in
a neural net.
\end{abstract}

\maketitle

\thispagestyle{fancy}

\section{Introduction}

At the Durham Statistics in Physics Conference (2002), S. Towers\cite{1} noted
some of the problems that occur when one uses many, correlated
variables in a multivariate analysis and proposed a heuristic method to shrink
the number.  In this note
a semi-automatic method is suggested help with this problem.  
The MiniBooNE experiment is faced with just such a problem,
distinguishing $\nu_e$ events from background events given a large
number of variables obtained from the event reconstructions.

\section{Fisher Discriminant Method and its Extension}

The Fisher discriminant method is a standard method for obtaining a
single variable to distinguish hypotheses starting from a large number of
variables.  If the initial variables come from a multi-normal
distribution, the Fisher variable encapsulates all of the
discrimination information.  However, in many problems the variables
are not of this form and the Fisher variable, although useful, is not
sufficient. 

The Fisher discriminant method \cite{2}
finds the linear combination $y$ of the
initial variables $x$ which maximizes 
\begin{equation*}
(\overline{y}_{\rm sig} - \overline{y}_{\rm bkg})^2/[{\rm var_{sig} +
var_{bkg}}],
\end{equation*}
where $\overline{y}$ is the mean value of the variable and var is the 
variance.  If $S$ is the
correlation matrix for the original variables corresponding to the 
denominator, then the inverse
of $S$ dotted into $(\overline{x}_{\rm sig} - \overline{x}_{\rm bkg})$, 
gives the
combination which maximizes the preceding expression.

If the distribution is not multi-normal, there is information still to
be obtained after finding the Fisher variable.
It is then useful to apply this method successively, firstly to the original
variables and, afterwards, to several non-linear transformations of the
variables.  Presently, three transformations are chosen: the
logarithms of the original variables, the exponentials of the original
variables and the cube of the original variables.  Together with the
original variables, this is then four choices.  The present note describes
a work in progress.  It is highly likely that
these are not optimum and that better choices can and will be found.
Indeed, it is likely that the optimum choice depends on the problem.

The method is also used with the original denominator, 
$[{\rm var_{sig} + var_{bkg}}]$ and then with each individual variance
in turn.  This is done since some of the variables may be quite narrow
for signal and wide for background or vice versa.  (See Figure 1)
There are then $4\times 3 = 12$ variables obtained with this method.
\begin{figure}[htbp]
\begin{center}
\includegraphics[width = 6cm]{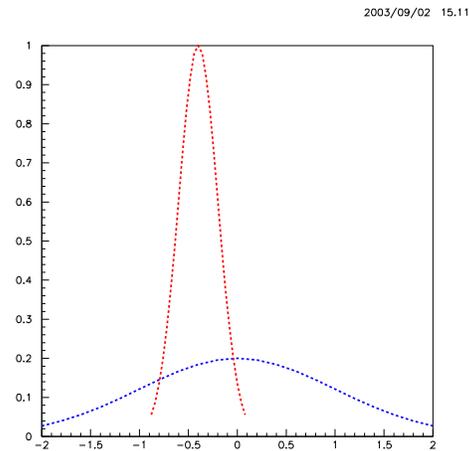}
\caption{Figure 1:  Different Width Normal Distributions}
\end{center}
\end{figure}

The procedure follows the following steps:
\begin{itemize}
\item{1.} Start with equal Monte Carlo samples of signal and
background events
\item{2.} Multiply and translate each variable  to have an overall mean
of zero and unit variance.  It is useful to fold variables if
necessary to maximize the difference in means.  A few events, very far
out on the tails of the distribution are clipped ($x>6\sigma$).
\item{3.} Order the variables according to 
$|\overline{x}_{\rm sig}-\overline{x}_{\rm bkg}|$ divided by the smallest of
the signal and background variances. At present, 
the ordering of variables is done only once.
\item{4.} Apply this extended Fisher method to the appropriate
transformation of the variables.
\item{5.} Use the Gram-Schmidt procedure to make the other variables
have zero correlation coefficient with the chosen linear combination.
\item{6.} The new variable is a linear combination of the original $n$
variables.  One variable must be discrded to have an independent set.
Discard the least significant (by the criterion of step 3) of
the original variables.  Using the $n-1$ non-Fisher variables, 
go back to step 2 to get the next variable.
\end{itemize}

For MiniBooNE the roefitter reconstruction started with 49 particle
identification variables.  Using the steps outlined, these were
reduced to 12 variables.  When $\nu_e$ quasi-elastic events were
compared with background neutral current $\pi^0$ events, the use of this
procedure with a neural net kept 46\% of the $\nu_e$ and reduced the
$\pi^0/\nu_e$ ratio to 1.1\% of its original value.  The neural net
was not hard to tune.  The
reconstruction--particle identification package is still being
improved, so these numbers will improve further.  The results obtained
here are similar to those obtained using a more elaborate neural net on
a subsample of 26 of the original 49 variables. 

These 12 variables have zero correlation coefficients.  Use of the neural 
net is simplified and it is convenient to look at the effect of cuts using
these variables.

Plots of the first nine of the twelve variables are shown in Figure 2.

\begin{figure*}[t]
\centering
\includegraphics[width = 15cm]{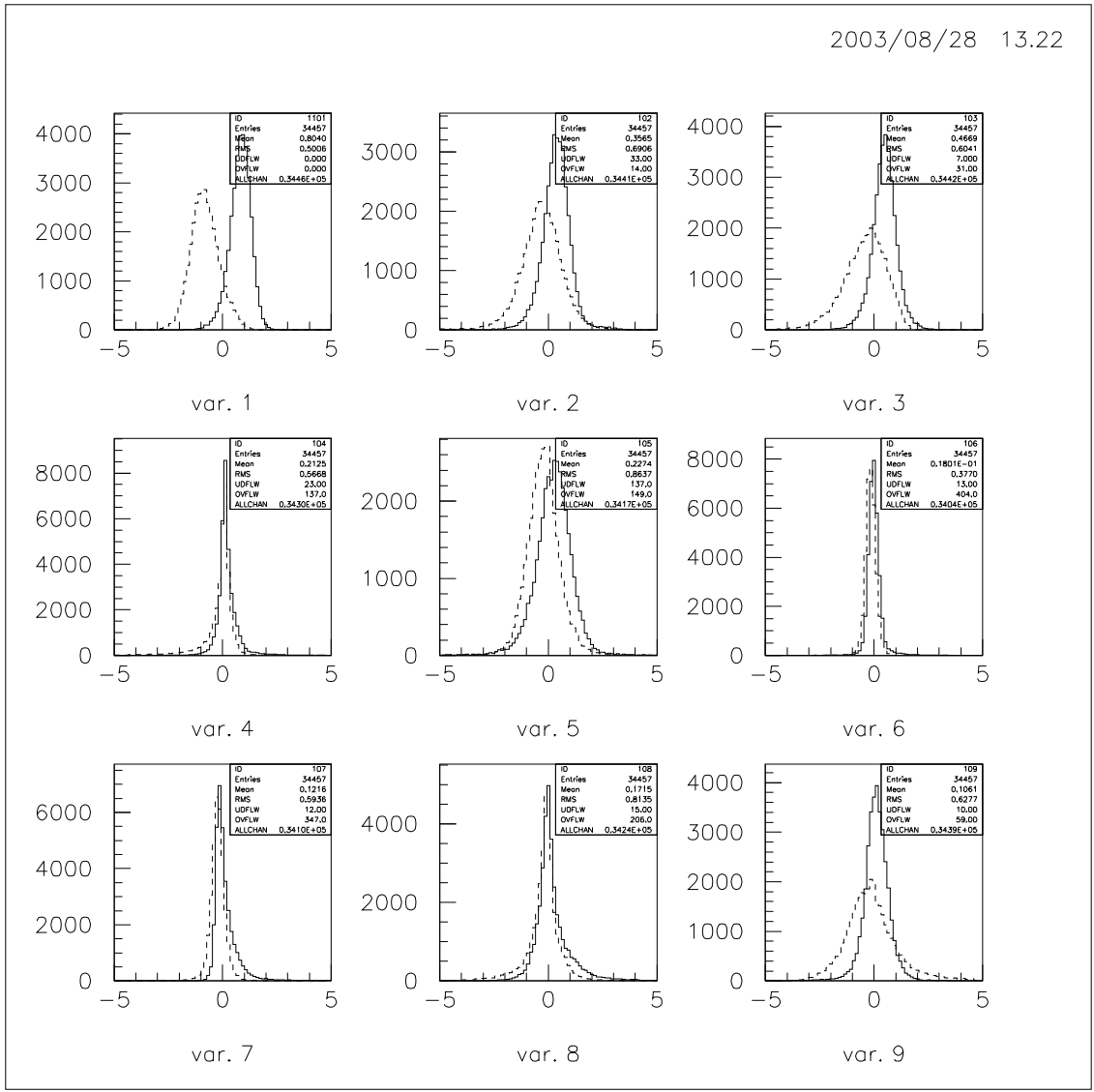}
\caption{Figure 2:  Plots of the first nine of the variables obtained.  The
solid lines are the $\nu_e$ signal and the dashed lines are the
$\pi^0$ background.}
\end{figure*}

\end{document}